# Modeling and Prediction of Iran's Steel Consumption Based on Economic Activity Using Support Vector Machines

Hossein Kamalzadeh[1a], Saeid Nassim Sobhan[b], Azam Boskabadi[c], Mohsen Hatami[d], Amin Gharehyakheh[e]

[a] Department of EMIS, Lyle School of Engineering, Southern Methodist University, E-mail: hkamalzadeh@smu.edu
[b] Department of Industrial Engineering, Amirkabir University of Technology, s.n.sobhan@aut.ac.ir
[c] Department of Industrial, Manufacturing, and Systems Engineering, University of Texas at Arlington, Arlington, Texas, USA, E-mail: azam.boskabadi@uta.edu
[d] M. E. Rinker, Sr. School of Construction Management, University of Florida, Gainesville, FL, USA, E-mail: mohsen.hatami@ufl.edu
[e] Department of Industrial, Manufacturing, and Systems Engineering, University of Texas at Arlington, Arlington, Texas, USA, E-mail: amin.gharehyakheh@uta.edu

## Abstract

The steel industry has great impacts on the economy and the environment of both developed and underdeveloped countries. The importance of this industry and these impacts have led many researchers to investigate the relationship between a country's steel consumption and its economic activity resulting in the so-called intensity of use model. This paper investigates the validity of the intensity of use model for the case of Iran's steel consumption and extends this hypothesis by using the indexes of economic activity to model the steel consumption. We use the proposed model to train support vector machines and predict the future values for Iran's steel consumption. The paper provides detailed correlation tests for the factors used in the model to check for their relationships with the steel consumption. The results indicate that Iran's steel consumption is strongly correlated with its economic activity following the same pattern as the economy has been in the last four decades.

## 1 Introduction

The steel industry has many impacts on the development of a country and its environment. The steel industry produces a great amount of $CO_2$ emissions which has very harmful effects on the environment of the earth [1], [2]. Steel is used for hard infrastructural building of countries, in a vast variety of industrial and household products, and in machinery and tools. According to the trends, since in the last few decades many countries tried to rebuild their infrastructure and economy after the vast destructions of the Second World War, the total metal consumption in the world including steel has increased significantly, from about 500 million tons in 1967 to about one and half billion tons in 2014 [3], [4]. Due to this increase in steel production and consumption many researchers have recently studied its trends and the factors causing this intensification in steel consumption in different countries such as China [5]–[7], United Kingdom's [8], [9], Korea [10], Poland [11], India [12], Japan [13], United States [14] and Australia [15]. As well as these trends being analyzed in each country individually, global steel demand

---

[1] Corresponding author: Department of EMIS, Lyle School of Engineering, Southern Methodist University, P.O. Box 750123, Dallas, TX 75275-0123, email: hkamalzadeh@smu.edu, Tel: +1 (213) 292 1746

has also been analyzed and conclusions have been made upon it [1]. Although many steel producing and consuming countries have been studied lately, some countries including Iran, have been disregarded and never studied.

Iran is one of the largest steel producers in the Middle East [3] with the annual steel production of about 25 million tons in 2018 (compare it to China, the world's largest producer and consumer, with 928 million tons per year) and the annual steel consumption of about 20 million tons in 2017 (compare it to China with 736 million tons per year). Iran has the goal of producing 55 million tons of steel by 2025 while it produced only 25 million tons in 2018. The most important fact beside these is that Iran is a developing country which has recently (about 4 decades) survived from a revolution and an 8-years of difficult war. During these years, Iran has been re-establishing and re-building its infrastructure. The steel production industry of Iran has an age about 4 decades and it is considered to be a young industry. In this paper we will focus on this country and its economy and see how these facts all contribute to the better understanding of the so-called intensity of use hypothesis first formulated and introduced by Malenbaum in 1978 [16]. According to the intensity of use hypothesis, there seems to be a relationship between steel consumption and economic activity of a country [8], [10], [11]. In another word, since metal intensive sectors of a country such as transportation, construction and manufacturing, consume steel in great proportions, the economic activity or the activity of these sectors has high impacts on the amount of steel consumed in the country [1], [8]. Although these factors may vary from country to country, the basis of this hypothesis stays the same. According to the intensity of use hypothesis, the consumption of steel in a country is a function of the development of its industry, income of its population and the demand for the products produced by steel in that country [1].

The validity of the intensity of use hypothesis has been tested for many countries in various studies. It is proved that the accuracy of the intensity of use hypothesis or the existence of relation between steel consumption and economic activity depends on characteristics of each country and the historical records (such as political and economic fluctuations or sudden cataclysms like wars or evolutions) of the period for which the hypothesis is being validated.

For the United Kingdom's, a long time stationary equilibrium relationship between economic activity and steel consumption is examined in [8] based on the intensity of use hypothesis with fractionally integrated and cointegrated processes which enable the possibility of fraction in the equilibrium at occasional points in time. The results of this analysis proved that such a long term relationship exists for UK. For Korea also this relationship both in the form of long term and short term is checked in [10] using vector error correction and vector autoregression models based on the intensity of use hypothesis. The study proves that a long term equilibrium relationship exists between GDP and total steel consumption. For Poland, a customized model of intensity of use is developed in [11] and based on that, using regression, forecasts for steel consumption for five years were made. For China, a model of steel consumption is developed based on four major factors such as saturation levels, lifetime distributions, GDP and urbanization rate in [5] and it is proved that these factors affect steel demands both in long term and short term. Based on this model, estimations for the changes in steel demands were made. Also in [6] China's demand for steel importation is modeled by a Cointegration procedure and the results show that demand for importation of steel products which is a subset of total steel demand is strongly correlated with China's economic activity. For India Cointegration and Granger causality between steel consumption and economic growth is examined in a bivariate vector autoregression format in [12] and it is shown that although there is no Cointegration, there is a unidirectional Granger causality from economic growth to steel consumption, the fact that has been proved in most of the papers discussed above. For Japan also the intensity of use model is utilized to forecast steel consumption in Japan in [7] where authors consider six steel consuming industries as the main industries affecting the steel demand in Japan. Based on the forecasts made for each of these industries, an aggregate forecast is made for total steel consumption in Japan. For the United States and Australia the same analyses are performed respectively in [14] and [15] based on the intensity of use model.

According to these studies, it is deduced that the intensity of use hypothesis is not always a rule of thumb. We may not be able to apply it to every country for every period of time. Investigating such a hypothesis, a country-specific model should first be developed and then the hypothesis gets tested given the developed model. In this paper we first test the validity of the intensity of use hypothesis for Iran, then we develop a model for Iran's steel consumption based on its economic activity. Finally, we predict Iran's steel consumption for the next couple of years by using the proposed model.

This paper continues as follows. In the next section, we provide a description of support vector machines and how they perform the task of classification or regression. Then in section 3, we review the so-called intensity of use model and we check its validity for the case of Iran's steel consumption. Then section 4, describes our proposed model for Iran's steel consumption and in section 5, we use the developed model to train support vector machines to predict future values for Iran's steel consumption. Finally, section 6 concludes the paper.

## 2 Support Vector Machines

Support Vector Machines (also called Support Vector Networks or SVMs) in the form which is used nowadays were first invented by Vladimir N. Vapnik and Corinna Cortes in 1995 [17]. Although the main goal and motivation of this method was to perform a linear classification [18]–[20], later it was used also to do a non-linear classification, regression analysis [21]–[25], clustering [26]–[28], and prediction [29], [30], [39], [40], [31]–[38]. This method always provides the global optimum solution while being robust. The method results in a programming problem called quadratic programming which is solvable without using other methods while the dimension of the problem is not very high.

To classify the data points, support vector machines construct a hyperplane or a set of hyperplanes with the dimension equal to the dimension of the problem. The goal is to achieve the best or optimal plane $w^T\varphi(x_i) + b$ that has the largest distance or margin from the nearest data points. Sometimes the data points are not linearly separable and thus it is needed to use a non-linear separator. Consider $D = \{x_i, y_i\}_{i=1}^N$ where $N$ is the total number of datapoints, $x_i$ is the input datapoints and $y_i \in \{1, -1\}$ is the corresponding class of each datapoint. The SVM has the following optimization problem adapted form [28], [29]:

$$minimize(w, b, \varepsilon_i) \quad \frac{1}{2}w^T w + C \sum_i \varepsilon_i \quad (1)$$

$$subject\ to \quad y_i(w^T\varphi(x_i) + b) \geq 1 - \varepsilon_i \qquad \varepsilon_i \geq 0, i = 1, \dots, N,$$

where $\varphi(.)$ is the mapping to the high dimensional feature space, $C > 0$ is the parameter controlling the tradeoff between training errors and model complexity, $\varepsilon_i$ are the slack variables. To solve the problem a dual problem is formulated with the use of Lagrange multipliers $\alpha_i$.

$$minimize(\alpha_i) \quad -\frac{1}{2}\sum_{i,j} \alpha_i \alpha_j y_i y_j k(x_i, x_j) + \sum_i \alpha_i \quad (2)$$

$$subject\ to \quad \sum_i \alpha_i y_i = 0, \qquad 0 \leq \alpha_i \leq C, i = 1, \dots, N$$

where $k$ is the kernel function $k(x_i, x_j) = \varphi(x_i)^T \varphi(x_j)$ for which in this paper Gaussian/RBF Kernel function is selected as shown below:

$$k(x_i, x_j) = \exp\left(-\frac{\|x_i - x_j\|^2}{\sigma^2}\right). \quad (3)$$

Solving the dual quadratic programming problem, results in a decision function for each datapoint as follows:

$$f(x) = w^T \varphi(x) + b = \sum_{i=i}^N \alpha_i y_i k(x_i, x) + b = \sum_{i \in SV} \alpha_i y_i k(x_i, x) + b. \quad (4)$$

Support vectors are those datapoints for which $\alpha_i$ is nonzero. Thus, $sign(f(x))$ simply defines the class of each datapoint. But this type of SVM only works when the problem is a classification type and the classes are nominal. In our specific problem we are trying to predict a continuous real value using predictors. In this case we need to use the regression type of SVMs called SVR or support vector regression. The principles of this method are completely similar to SVMs while SVR finds the optimal trend line for a given set of datapoints.

The parameter $C$ in SVM or SVR needs to be given a value. Thus, SVR should be tuned with respect to this parameter. We will do so in section 5 of the paper.

## 3 Intensity of Use Model for Iran's Steel Consumption

The economic activity and consumption of steel of a country are tightly related to each other [10], [11], [16]. Based on this relationship, we can model a country's steel consumption as a function of specific measures of its economic activity [8], [10]. The intensity of use hypothesis was first formulated by Malenbaum in [16]. According to this hypothesis, the intensity of use is defined as the amount of steel consumed per a nation's output as formulated below:

$$IU_t = \frac{\text{Steel consumption in year } t}{\text{GDP in year } t}. \tag{5}$$

The relationship in Eq. (5) can also be represented in the following form [8]:

$$C_t = \delta Y_t^\lambda \quad \text{or} \quad IU_t = C_t/Y_t = \delta Y_t^{\lambda-1}, \tag{6}$$

where $C_t$ is the steel consumption in the year $t$, $Y_t$ is the output of the country in the year $t$ and $\delta$ and $\lambda$ are constants.

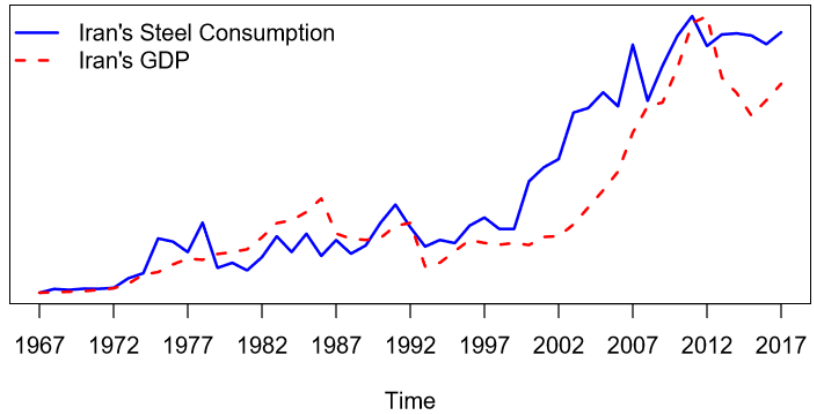

*Figure 1 Iran's steel consumption and GDP from 1967 to 2017*

To check whether this relationship in Eq. (6) exists between Iran' steel consumption and GDP (both plotted in Figure 1), we take the natural log of both sides of the equation to linearize the relationship between $C_t$ and $Y_t$. This results in:

$$\ln(C_t) = \ln(\delta) + \lambda \ln(Y_t). \tag{7}$$

The relationship between $C_t$ and $Y_t$ in Eq. (7) has now a linear form which makes it easier to be tested using the various tests. We apply linear regression on both time series $C_t$ and $Y_t$ to check for a meaningful relationship or correlation between these two variables and to see if the linear relationship in Eq. (7) exist between these two variables. Table 1 contains the results for the regression analysis on the Eq. (7).

*Table 1 Regerssion analysis results for the intensity of use model on Iran's steel consumption and GDP*

| $\lambda$ | Ln($\delta$) | RSE | R-square | Adjusted R-square | p-value |
|---|---|---|---|---|---|
| 0.82 | -0.045 | 0.463 | 0.8603 | 0.8574 | 2.2e-16 |

The results in Table 1 including the very small p-value indicate strong evidence against the null hypothesis, rejecting the hypothesis that the two variables are not related. Figure 2 also shows the linear regression between $\ln(C_t)$ and $\ln(Y_t)$ in Eq. (7) as well as the 95% confidence intervals.

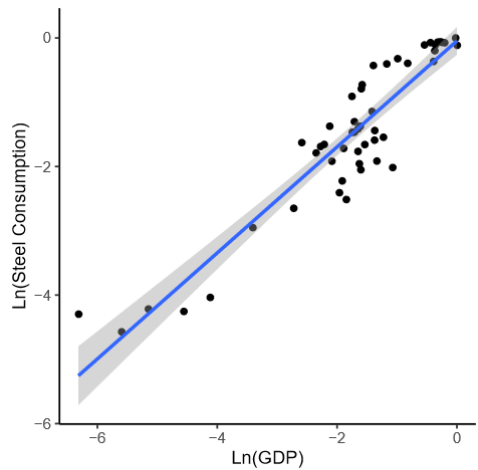

*Figure 2 Linear regression between Iran's Steel consumption and GDP (intensity of use model)*

Beside the results indicated in Table 1 and Figure 2, it is obviously clear from Figure 1 that these two trends have been changing with a similar pattern over the past 50 years. According to Figure 1, there exist some unusual changes both in steel consumption and GDP that interestingly happened to both of the trends at the same period of time, first in 1970s and second in 1980s. Looking back to the historical events happened in Iran, there are two distinct events that are of interest: the first in 1970s was the revolution resulted in a change in the governing regime and the second in 1980s was an 8-years devastative war, both hindering the country from developing and building the infrastructure, incurring huge costs on the economy and also requiring the country to rebuild some infrastructures in the attacked cities. The cause for the sudden reduction in both steel consumption and GDP about the end of the 1970s is the change of the regime as it can easily hinder many sections of the country's economy. The cause for the sudden increase both in steel consumption and GDP right at the end of the 1980s was the need for rebuilding infrastructure in the country and the need to produce more to compensate for what had happened in the previous decade.

## 4 Proposed Model for Iran's Steel Consumption

We already showed using regression analysis that a long-term relationship exists between Iran's steel consumption and economic activity in the last 50 years. It also seems that this relationship is going to last for some time while Iran is still building its infrastructure and developing. In this paper we focus on developing a model for the Iran's steel consumption that not only incorporates the intensity of use hypothesis but also takes into account other economic activities of the country. We propose our model based on the fact that steel is mostly consumed in specific industries and economic sectors of a country such as construction, shipbuilding, transportation and railways, petroleum, manufacturing such as machinery and automobiles, and home appliances and consumer durables, all metal intensive sectors of a country's economy [5], [8], [10]. These sectors of the economy are incorporated in our model by implementing the following list of economic indexes:

- GDP per capita (constant 2005 US$)
- GDP (constant 2005 US$)
- Manufacturing, value added (constant 2005 US$)
- Industry, value added (constant 2005 US$)
- Crude oil production (million barrels per year)
- Energy production (kilo tons of oil equivalent)
- Rail Lines (total route-km)
- Urban population

The largest common period of the mentioned indexes for which the data is available is from 1967 to 2017. Figure 3 to Figure 10 depicts each index individually along with steel consumption over this period (due to the different units, data has been standardized).

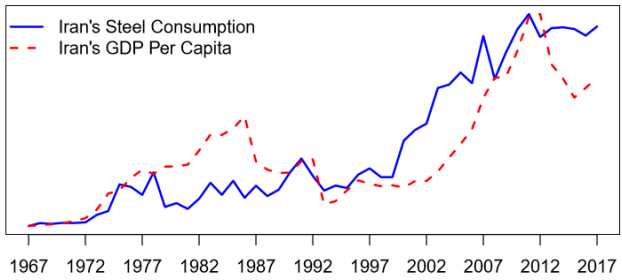
*Figure 3 Iran's GDP per capita and steel consumption*

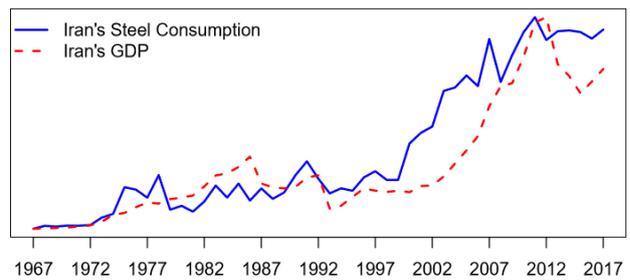
*Figure 4 Iran's GDP and steel consumption*

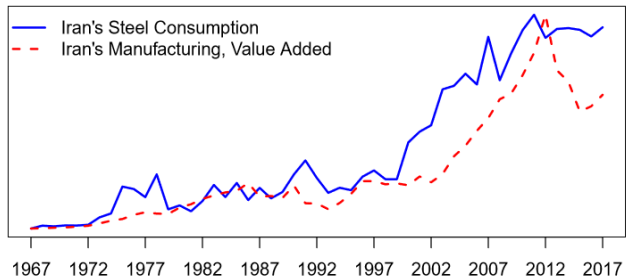
*Figure 5 Iran's manufacturing value added and steel consumption*

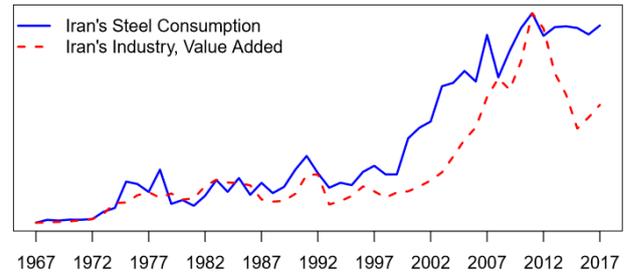
*Figure 6 Iran's industry value added and steel consumption*

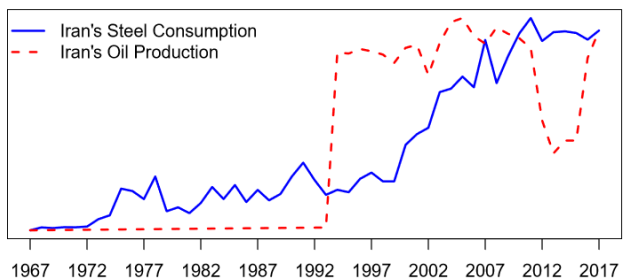
*Figure 7 Iran's crude oil production and steel consumption*

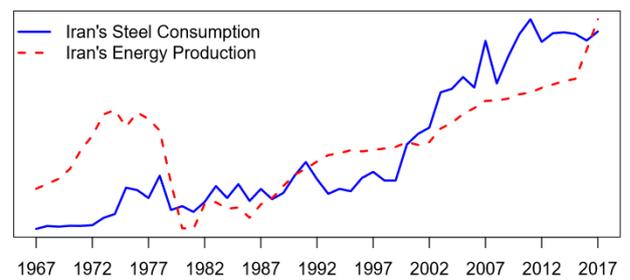
*Figure 8 Iran's energy production and steel consumption*

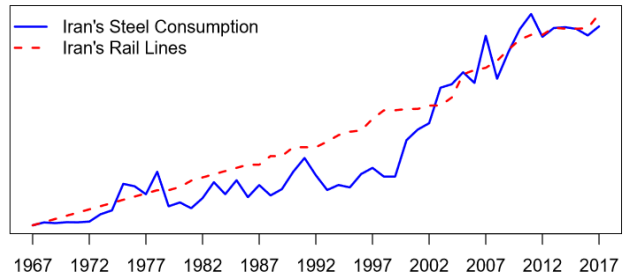
*Figure 9 Iran's rail lines and steel consumption*

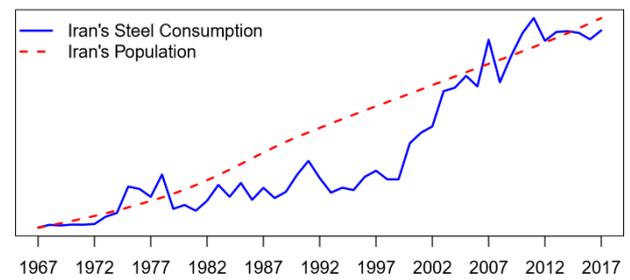
*Figure 10 Iran's urban population and steel consumption*

Similar to the case of Iran's steel consumption and GDP, there seems to be a relationship between each of these indexes and Iran's steel consumption. Majority of them seems to be changing with a similar pattern as the steel consumption's. To check this, we performed three different correlation analyses on each of the indexes. The results of these tests are provided in Table 2. The table provides correlation coefficient of each test along with the statistic of the test, and significance of it as provided under the p-value column. According to this column for all tests, the null hypothesis (no correlation between the two variables) is rejected and therefore a correlation exists between the variables.

*Table 2 Correlation analysis results for Iran's economic activity and steel consumption*

| Analysis | Pearson | | | Kendall | | | Spearman | | |
|---|---|---|---|---|---|---|---|---|---|
| Factor | coefficient | t | p-value | tau | z | p-value | rho | s | p-value |
| **GDP** | 0.9168 | 16.070 | 3.61E-21 | 0.6920 | 7.164 | 7.83E-13 | 0.8507 | 3298.57 | 2.71E-15 |
| **GDP per capita** | 0.8423 | 10.939 | 9.42E-15 | 0.5555 | 5.751 | 8.89E-09 | 0.7230 | 6120.64 | 2.07E-09 |
| **Manufacturing** | 0.9351 | 18.476 | 1.00E-23 | 0.7689 | 7.960 | 1.72E-15 | 0.9078 | 2036.55 | 3.97E-20 |
| **Industry** | 0.9145 | 15.823 | 6.84E-21 | 0.7250 | 7.505 | 6.13E-14 | 0.8847 | 2548.56 | 7.34E-18 |
| **Oil production** | 0.6919 | 6.709 | 1.88E-08 | 0.6444 | 6.669 | 2.58E-11 | 0.8198 | 3983.18 | 1.88E-13 |
| **Energy production** | 0.7681 | 8.397 | 4.76E-11 | 0.5555 | 5.751 | 8.89E-09 | 0.7079 | 6454.65 | 6.26E-09 |
| **Rail lines** | 0.9485 | 20.968 | 4.03E-26 | 0.8175 | 8.441 | 3.14E-17 | 0.9338 | 1463.80 | 1.65E-23 |
| **Urban population** | 0.9097 | 15.343 | 2.02E-16 | 0.8128 | 8.327 | 2.21E-16 | 0.9326 | 1488.5 | 2.21E-16 |

Among the indexes studied in this paper, according to the results of the tests, rail lines, manufacturing and industry sectors of the Iran's economy have the highest correlation with steel consumption. This was expected as these three sectors are the ones that highly demands steel in large volumes (metal intensive sectors). Rail lines are a major way of transportation and logistics in a country and also an indicator of how much the cities in a country are connected to each other. The results of all tests are almost consistent for these three indexes.

GDP is also highly correlated to steel consumption which contributes to the so-called intensity of use model studied earlier in the paper. The reason behind this can be the fact that GDP is a general index of economy accounting for many sectors of the economy of a country. Urban population also seems to be highly correlated with the steel industry. This is due to the fact that urbanization has a great impact on steel consumption in a country, i.e., the larger the urban population is the more need will be for the infrastructure in the cities. This demands more steel to be consumed. Iran is one of the largest producers of oil in world and in the middle east. This industry consumes steel not directly but for the building of its infrastructure. A large proportion of Iran's GDP is also coming from oil exportations. The energy production can also be a great indicator of steel consumption in a country specifically a country like Iran which is still under development. Building the infrastructure for producing energy such as power plants, dams, wind turbines and so on requires steel.

Table 2 clearly indicates that these indexes are correlated with Iran's steel consumption. We here state that not only can the GDP be used to model the steel consumption, but also other factors including the ones introduced in this section can be great indicators of steel consumption. These indexes are all economic-activity-indicators the same as GDP but for more specific sectors of the economy. In our steel consumption model, we incorporate all these factors at the same time each for the reasons mentioned above. In the next section of the paper, we use SVR to train this model and later predict future values for Iran's steel consumption using the trained model.

## 5 Iran's Steel Consumption Prediction Using the Proposed Model

Using the model proposed in the previous section that implements 8 indicators of steel consumption, we develop an SVR (support vector regression) model and train the SVR using the data available from 1967 to 2017. The data used for this purpose are all in the form of time series. These data are depicted in Figure 3 to Figure 10. The time series seem to be not stationary and have trends. We need to check for these characteristics in advance and make the necessary transformations to make the time series stationary and trendless.

### 5.1 Time series pre-processing

Time series data often requires some preprocessing prior to being modeled with machine learning algorithms [41]. Typical characteristics of this type of data is having trends, not being stationary, or having seasonality. Each of these characteristics should be removed to prior to processing, using appropriate transformations. The series shown in Figure 3 to Figure 10 seems to have trends. Also, the stationarity of these series needs to be checked using proper tests. We used Augmented Dickey-Fuller (ADF) test [42] to check for the stationarity of the time series. A p-value less than 0.05 indicates the time series are stationary. We also performed Kwiatkowski-Phillips-Schmidt-Shin (KPSS) test [43] to check for the stationarity of the series. It seems the results of KPSS tests are consistent with the ADF test. A p-value above 0.05 indicates that the time series is stationary in a KPSS test. **Table 3** contains the results for these two tests for each of the series. These results indicate that the respective time series are not stationary, and they need to be transformed by differencing.

*Table 3 Stationarity test results for the economic activity time series*

| Analysis | ADF test | | | KPSS test | | |
|---|---|---|---|---|---|---|
| Series | Parameter (lag order) | test statistic | p-value | Truncation lag parameter | test statistic | p-value |
| **Steel consumption** | 3 | -1.45793 | 0.79229 | 3 | 1.191154 | 0.01 |
| **GDP** | 3 | -1.77913 | 0.663165 | 3 | 0.985944 | 0.01 |
| **GDP per capita** | 3 | -1.98967 | 0.578524 | 3 | 0.802824 | 0.01 |
| **Manufacturing** | 3 | -1.87613 | 0.624171 | 3 | 1.07787 | 0.01 |
| **Industry** | 3 | -2.34265 | 0.436624 | 3 | 0.947151 | 0.01 |
| **Oil production** | 3 | -2.44514 | 0.395422 | 3 | 1.039756 | 0.01 |
| **Energy production** | 3 | -1.16471 | 0.903951 | 3 | 0.754679 | 0.01 |
| **Rail lines** | 3 | -1.3051 | 0.853729 | 3 | 1.346072 | 0.01 |
| **Urban population** | 3 | -3.37277 | 0.06988 | 3 | 1.375549 | 0.01 |

We also performed Canova-Hansen test [44] and Osborn-Chui-Smith-Birchenhall test [45] for seasonality and their results indicate that the series are not seasonal.

In order to make the time series stationary, we perform differencing on each one. Differencing is to subtract each data point in the series from its successor [41]. The minimum number of differences required to make the series stationary is returned by the ADF test. The result for just the steel consumption series is depicted in **Figure 11** compared to the original time series.

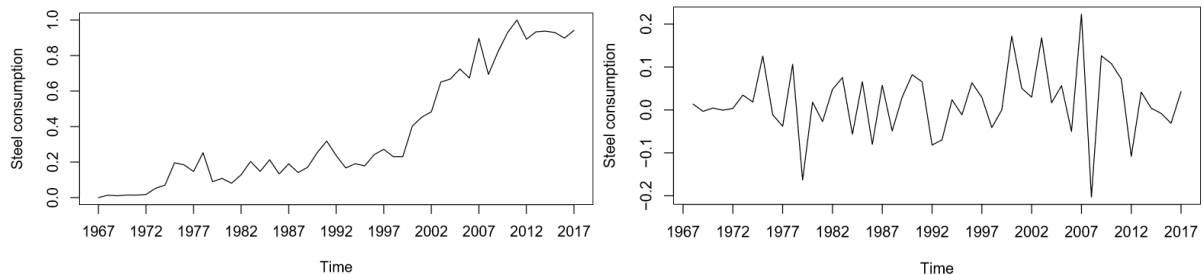

*Figure 11 Iran's steel consumption time series before and after differencing and detrending*

Although this paper is basically dealing with lots of time series as the input data, and lots of preprocessing techniques are developed and available in the relative literature [46]–[53], digging into these techniques and implementing them for such short time series as in this paper are avoided due to short lengths of the times series used.

## 5.2 Predicting using SVR

In order to predict using SVR we first need to train the SVR using part of the data. We randomly partition 70 percent of the data into train set and the rest into the test set. We then use the train set to train the SVR with a linear kernel. The training is done using repeated cross validation with 10 folds repeated 3 times with parameter $C = 1$. The training results in an RMSE of 0.0668, R-squared of 0.942, and MAE of 0.0576. The testing using the same model results in an RMSE of 0.119.

In order to tune the SVR with respect to parameter $C$, we train the SVR using a range of values for $C$. The results indicate that the best value for $C$ is 1.5 as demonstrated in **Figure 12** since it has the lowest RMSE among the rest. Using $C = 1.5$ we test the SVR again and the testing results in an RMSE of 0.118 which is slightly lower than using $C = 1$. The comparison between the actual values and predicted values for the test set is demonstrated in **Figure 13**.

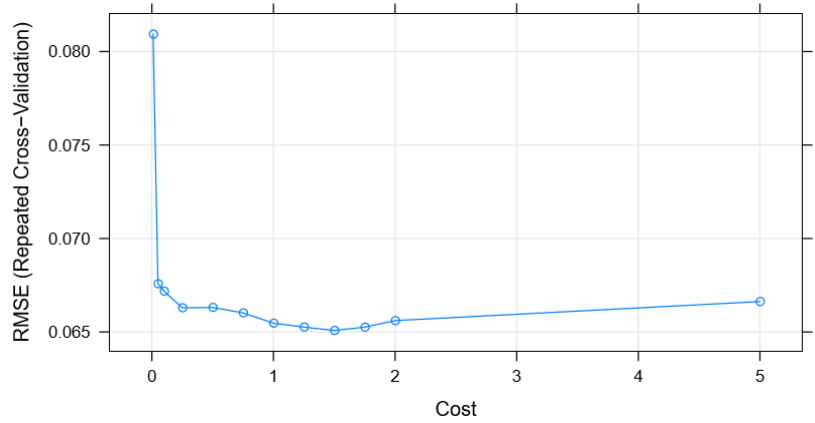

*Figure 12 RMSE values as a function of parameter C*

We can use the final SVR model to predict future values for the steel consumption. We estimate future values for each of the eight predictors using linear regression and then use the estimated values to predict the future values for the case of steel consumption. The predictions for the next 10 years starting from 2018 are provided in Table 4.

*Table 4 Predictions for Iran's steel consumption for the next 10 years in million barrels*

| Year | Prediction (mb) |
|---|---|
| 2018 | 20.491 |
| 2019 | 20.922 |
| 2020 | 21.352 |
| 2021 | 21.783 |
| 2022 | 22.214 |
| 2023 | 22.645 |
| 2024 | 23.076 |
| 2025 | 23.507 |
| 2026 | 23.938 |
| 2027 | 24.369 |

The predictions in Table 4 show that Iran's steel consumption will not be increasing rapidly. This can be an indicator of a slow growth in Iran's economy which could be due to the economic sanctions. This affects the rate of infrastructure building and improvement, slowing it down by a great factor and consequently decreasing the rate of steel consumption in the country.

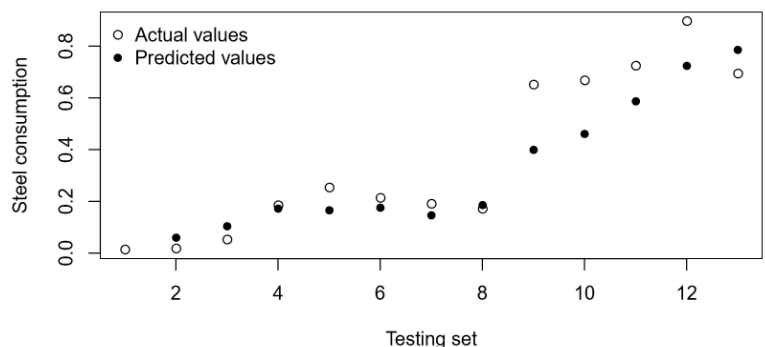

*Figure 13 Actual values and predicted values for Iran's steel consumption using the tuned SVR (test set)*

# 6  Conclusion

We investigated the validity of the intensity of use hypothesis for the case of Iran's steel consumption. Based on the intensity of use hypothesis, a relationship exists between a country's steel production and economic activity. GDP is one of the most common indicators of a nations economic activity. Based on the intensity of use hypothesis, steel consumption can be strongly related to the GDP. We investigated this by formulating the hypothesis for Iran's steel consumption and checked if these two are related or not. The results of the regression analysis of these two variables indicate that these two are strongly related to each other. Thus, the intensity of use hypothesis seems to hold for the case of Iran's steel consumption. Given this, we expanded the model in a way that includes more indicators of a country's economic activity by incorporating indexes such as manufacturing value added, industry value added, oil production, energy production, and rail lines. We checked the correlation between each of these factors and Iran's steel consumption, and results of the tests indicate that all these factors are strongly correlated to steel consumption. We developed a model using all these factors as predictors for the Iran's steel consumption and trained the model with SVR, a specific kind of SVMs. The tuned SVR can predict the future values for Iran's steel consumption using the proposed model.

The steel consumption and economic activity seems to have a complex behavior in countries that are still developing including Iran. In these countries, infrastructure is still being built and steel is being consumed for that. But for the case of Iran, it is different as Iran has experienced tough situation during the past 4 decades, a revolution resulting in a change in regime, an eight-year war that devastated the country's economy at the end, and recently economic sanctions. These incidents have had high impacts on the economy of the country, and this is noticeable in the trends of economic activity. These sudden huge changes make it difficult to understand the hidden patterns of the Iran's steel consumption though their impact on the economy and steel consumption is very clear. Iran's economy is suffering these days from many sanctions and this affects the steel consumption trend and other economic activities of the country making it very difficult to predict what happens to steel consumption in this country. We predict that Iran's steel consumption will suffer a slow-down in the years coming and despite the fact that it is increasing, the rate of this increase would gradually decrease until the steel consumption decreases in the next decade. If the economy of the country stays in the same way, the steel consumption would most probably go down.

Future studies can investigate the exact effects of these sanctions on the economic activities of this country as well as its steel consumption. Future studies can focus on incorporating more economic factors that implement the effects of the sanctions in the economy. Also, since the energy price for the industry section of the country is cheaper compared to many other developed countries in the world, we believe that the production of steel might also affect its consumption in Iran. This can also be investigated in future studies. Future studies can also focus on using other methods of prediction and also on how to weight each factors in steel consumption model.